\documentclass[10pt,twocolumn]{article}
\usepackage[a4paper, total={7in, 7.9in}]{geometry}
\usepackage{graphicx}
\usepackage{booktabs}
\usepackage{multirow}
\usepackage{float}
\usepackage{multicol}
\restylefloat{table}

\title{Affect Enriched Word Embeddings for News Information Retrieval}

\author{
Tommaso Teofili \\ Adobe \\ teofili@adobe.com
\and
Niyati Chhaya \\ Adobe \\ nchhaya@adobe.com
}

\begin{document}
\date{}
\maketitle

\begin{abstract}
Distributed representations of words have shown to be useful to improve the effectiveness of IR systems in many sub-tasks like query expansion, retrieval and ranking. Algorithms like word2vec, GloVe and others are also key factors in many improvements in different NLP tasks. One common issue with such embedding models is that words like \textit{happy} and \textit{sad} appear in similar contexts and hence are wrongly clustered close in the embedding space.
In this paper we leverage Aff2Vec, a set of word embeddings models which include affect information, in order to better capture the affect aspect in news text to achieve better results in information retrieval tasks, also such embeddings are less hit by the synonym/antonym issue. 
We evaluate their effectiveness on two IR related tasks (query expansion and ranking) over the New York Times dataset (TREC-core '17) comparing them against other word embeddings based models and classic ranking models.
\end{abstract}

\section{Introduction}
Distributed representations of words, also known as word embeddings, have played a key role in various downstream NLP tasks. Such vector representations place vectors of semantically similar words close in the embedding space, allowing for efficient and effective estimation of word similarity.
Word2vec \cite{mikolov2013efficient} and GloVe \cite{pennington2014glove} are among the most widely adopted word embedding models because of their effectiveness in capturing word semantics. One of the advantage of using word embeddings in information retrieval is that they are more effective in capturing query intent and document topics than other local vector representations traditionally used in IR (like TF-iDF vectors).
Text tokens in IR don't always overlap with exact words; tokens often coincide with subwords (e.g. generated by stemmers), ngrams, shingles, etc. Therefore word embeddings are also often referred to as \textit{term} embeddings in the context of IR. Term embeddings can be used to rank queries and documents; in such context a dense vector representation for the query is derived and scored against corresponding dense vector representations for documents in the IR system. Query and document vector representations are generated by aggregating term or word embeddings associated with their respective text terms from the query and document texts.
Word embeddings can also be used in the query expansion task. Term embeddings are used in such contexts to find good expansion candidates from a global vocabulary of terms (by comparing word vectors), such enriched queries are used to retrieve the documents.
Most of recent good performing word embedding models are generated in an unsupervised manner by learning word representations looking at their surrounding contexts. However one issue with word embeddings is that words with about opposite meanings can have very similar contexts, so that, for example, `happy' and `sad' may lie closer than they should be in the embedding space, see related efforts in \cite{chen2015revisiting} and \cite{nguyen2016integrating}.
In order to mitigate this semantic understanding issue, we propose to use affect-enriched word embedding models (also known as Aff2Vec\cite{khosla2018aff2vec}) for IR tasks, as they outperform baseline word embedding models on word-similarity task and sentiment analysis.
Our contribution is the usage of Aff2Vec models as term embeddings for information retrieval in the news domain.
Beyond the synonym-antonym issue we except Aff2Vec models to work well for news IR because of their capability of better capturing writers' affective attitude towards articles' text (see section \ref{sec:aff-score}).
We present experiments against standard IR datasets, empirically establishing the utility of the proposed approach.

\subsection{Affect scores in news datasets}
\label{sec:aff-score}

In order to assess the potential applicability of Aff2Vec embeddings in the context of information retrieval, we run preliminary evaluation of the amount of formality, politeness and frustration contained in common text collections used in information retrieval experiments.
For this purpose we leverage the \textit{affect scoring} algorithm that is used for building Aff2Vec embeddings.
We extract mean affect scores for formality, politeness and frustration on each dataset.
Such an evaluation involves two collection of news: the datasets from TREC core 2018 track, Washington Post articles, and TREC core 2017 track, New York Times articles. Also we extract affect scores from the ClueWeb09 dataset \cite{callan2009clueweb09}, containing text of HTML pages crawled from the Web, and the CACM dataset, a collection of titles and abstracts from the CACM journal.
Results are reported in table \ref{tab:affect-mean}.

\begin{table}[H]
\begin{tabular}{llll}
\hline
\multicolumn{4}{c}{Dataset affect scoring} \\
\cline{1-4}
Dataset & formality & politeness & frustration \\
\hline
NYT & 0.7087 & 0.6291 & 0.6248 \\
WP & 0.7788 & 0.7456 & 0.6510 \\
CACM & 0.3619 & 0.1229 & 0.3511 \\
ClueWeb09 & 0.4319 & 0.2708 & 0.6216 \\
\hline
\end{tabular}
\caption{Mean affect scores on some common IR datasets}
\label{tab:affect-mean}
\end{table}

The scores for formality, politeness and frustration extracted on the Ney Work Times and Washington Post articles are generally higher than the ones extracted for CACM and ClueWeb09 datasets, except for the frustration score reported for ClueWeb which is very close to the frustration score extracted for NYT articles.
These results suggest that Aff2Vec embeddings should work well on the news domain as they are built to appropriately capture such affective aspects of information.

\section{Related work}

Dict2vec\cite{D17-1024} builds word embeddings using online dictionaries and optimizing an objective function where each word embedding is built via positive sampling of strongly correlated words and negative sampling of weak correlated ones \cite{tissier2017dict2vec}.
In \cite{zamani2017relevance}, embeddings are optimized using different objective functions in a supervised manner based on lists of queries and related relevant and non-relevant results.
In \cite{fornander2016generating}, word vectors in combination with bilingual dictionaries are used to extract synonyms so that they can be used to expand queries.
Documents are represented as bags of vectors generated as mixture of distributions in \cite{roy2016using}.
Efforts like \cite{chen2015revisiting} and \cite{nguyen2016integrating} are related to our work in the fact that they can be incorporated in usage of term embeddings in IR tasks.
For our ranking scenario, \cite{roy2016representing} is relevant as documents and queries are represented by mixtures of Gaussians over word embeddings, each of the Gaussians centered around centroid learned via e.g. a k-means algorithm. The likelihood of a query with respect to a document is measured by the distance of the query vector from each centroid that document belongs to, using \em{centroid similarity} or \em{average inter-similarity}.

\subsection{Aff2Vec:~Affect-enriched embeddings~\cite{khosla2018aff2vec}}
Word representations historically have only captured semantic or contextual information, but ignored other subtle word relationships such as difference in sentiment. Affect refers to the feeling of an emotion or a feeling \cite{Picard1997}. Words such as `glad', `awesome', `happy', `disgust' or `sad' can be referred to as affective words. Aff2Vec introduces a post-training approach that introduces `emotion'-sensitivity or affect information in word embeddings. Aff2Vec leverages existing affect lexicon such as Warriner's lexicon \cite{warriner2013norms} which has a list of over 14,000 English words tagged with valence (V), arousal (A), and dominance (D) scores. The affect-enriched embeddings introduced by Aff2Vec are either built on top of vanilla word embeddings i.e. word2vec, GloVe, or paragram or introduced along with counterfitting \cite{mrkvsic2016counter} or retrofitting \cite{faruqui2014retrofitting}. In this work, we leverage these enriched vector spaces too in order to evaluate their performance for standard IR tasks, namely - query expansion and ranking.

\section{Word embeddings for query expansion}

We leverage word embeddings to perform query expansion in a way similar to \cite{roy2016using}. For each query term $q$ contained in the query text $Q$, the word embedding model is used to fetch $w_q$ nearest neighbour $w_e$ in the embedding space, so that $cos(w_e,w_q) > t$, where $t$ is the minimum allowed cosine similarity between two embeddings to consider the word $e$ associated to the vector $w_e$ a good expansion for the word $q$ associated with the query term vector $w_q$.
Upon successful retrieval of an expansion of at least a term $q$ in a query, a new "alternative" query $A$ where $q$ is substituted by $e$ is created. Consequently the query to be executed on the IR system becomes a \textit{boolean} query of the form \textit{Q OR A}.
If more than one query term has a valid expansion fetched from the embedding model, all possible combinations of query terms and relative expansion terms is generated.
For example, given a query "recent research about AI", if term embeddings output that $nearest(recent) = latest$ with $cos(recent, latest) = 0.8$ bigger than the threshold $0.75$, the output query will be composed by two \textit{optional} clauses: \textit{"recent research about AI" OR "latest research about AI"}.

\section{Word embeddings for ranking}
In order to use word embedding models for ranking we chose to use the averaging word embeddings approach (also known as \textit{AWE}). Each document and query vector is calculated by averaging the word vectors related to each word in documents and query texts. The query / document score is measured by the cosine similarity between their respective averaged vectors, as in other research works like \cite{mitra2016dual, roy2016representing, rekabsaz2017toward, DBLP:conf/gi/GalkeSS17}.
In our experiments we used each word TF-iDF vector to normalize (divide) the averaged word embedding for query and document vectors. We observed that using this technique to smooth the sum of the word vectors instead of just dividing it by the number of its words (mean) resulted in better ranking results. 
This seems in line with the findings from \cite{schnabel2015evaluation} which indicate that cosine similarity may be polluted by term frequencies when comparing word embeddings.

\section{Experiments}

We compare the usage of Aff2Vec word embeddings in the ranking and query expansion task against both \textit{vanilla} embedding models (like word2vec and GloVe) and enriched models like Dict2vec models \cite{tissier2017dict2vec}. We also present experiments with variants in Aff2Vec: counterfitted and retrofitted models with enriched affect information.
All the models used in our experiments are pretrained.
To setup our evaluations we use two open source toolkits \textit{Anserini} \cite{yang2017anserini} and \textit{Lucene4IR} \cite{azzopardi2017lucene4ir}, both based on Apache Lucene \cite{bialecki2012apache}.
We run ranking and query expansion experiments on the New York Times articles from the TREC Core '17 track \cite{allan2017trec} since it's a relevant dataset for the news domain.
For the sake of generalizability, we also conduct the same evaluations over the CACM dataset \cite{fox1983characterization}, a "classic" dataset for IR.
For the case of query expansion we include evaluation using WordNet \cite{miller1995wordnet} in order to provide an expansion baseline not based on word embeddings.

\subsection{Results}

Table \ref{tab:rankingNYT} shows performance for ranking experiments on the NYT dataset using different embeddings. 
We observe that usage of term embeddings doesn't give benefits in many cases, classic BM25 and query likelihood retrieval models provide better NDCG than almost all the models except the affect enriched ones. A GloVe retrofitted affect enriched embedding model is the top performing one for NDCG measure. On the other hand none of the term embedding ranking could outperform BM25 on the mean average precision measure.

\begin{table}
\begin{tabular}{lll}
\hline
\multicolumn{3}{c}{Ranking experiments on NYT} \\
\cline{1-3}
Model & NDCG & MAP \\
\hline
BM25 &  0.4334 & \textbf{0.1977}\\
QL & 0.4325 & 0.1913\\
\hline
\multicolumn{3}{c}{NON ENRICHED MODELS }\\
\hline
GloVe  &  0.4292 &  0.1883\\
GloVe.42B.300d & 0.4003 & 0.1690\\
GloVe.6B.100d  &  0.4291 & 0.1911\\
GloVe.6B.200d  &  0.4314 &  0.1964\\
GloVe.6B.300d  &  0.4316 &  0.1946\\
GloVe.6B.50d  &  0.4078 &  0.1760\\
GloVe-Twitter-100  &  0.4212 &  0.1849\\
GloVe-Twitter-200  &  0.4242 &  0.1873\\
GloVe-Twitter-50 & 0.4128 & 0.1798\\
GloVe-Twitter-25  &  0.3541 &  0.1377\\
w2v-GoogleNews-300 & 0.4294 & 0.1922 \\
dict2vec-dim100 & 0.4101 &  0.1885 \\
dict2vec-dim200 & 0.4155 & 0.1891 \\
dict2vec-dim300 & 0.4151 & 0.1899 \\
\hline
\multicolumn{3}{c}{ENRICHED MODELS }\\
\hline
counterfit-GloVe  &  0.3980 &  0.1720\\
GloVe-retrofitted  &  0.4216 &  0.1861\\
paragram-counterfit  &  0.3840 &  0.1580\\
paragram-74627  &  0.4337 &  0.1937\\
paragram-retrofitted  &  0.3969 &  0.1703\\
paragram-retrofitted-74627  &  0.3963 &  0.1698\\
w2v-76427  &  0.4328 &  0.1969\\
w2v-counterfit-header  &  0.3972 &  0.1721\\
w2v-retrofitted  &  0.4341 &  0.1914\\
\hline
\multicolumn{3}{c}{AFFECT ENRICHED MODELS }\\
\hline
counterfit-GloVe-affect  &  0.4311 &  0.1753\\
GloVe-affect &  0.4594 &  0.1926 \\
GloVe-retrofitted-affect-555  &  \textbf{0.4693} &  0.1948\\
paragram-affect  &  0.4619 &  0.1969\\
paragram-counterfit-affect  &  0.4339 &  0.1788\\
w2v-affect  &  0.4592 &  0.1926\\
w2v-counterfit-affect  &  0.4309 &  0.1766\\
w2v-retrofitted-affect  &  0.4601 &  0.1911\\

\hline
\end{tabular}
\caption{Ranking experiments on NYT}
\label{tab:rankingNYT}
\end{table}

Table \ref{tab:expansionNYT} shows performance for query expansion experiments on the NYT dataset using different embeddings. 
We observe that classic BM25 and query likelihood retrieval models provide better NDCG than almost all the models except some of the affect enriched ones. This is in line with what we observed for the ranking task on the same dataset.
A GloVe retrofitted affect enriched embedding model is the top performing one for both NDCG and MAP. 

\begin{table}[H]
\begin{tabular}{llr}
\hline
\multicolumn{3}{c}{Query expansion experiments on NYT} \\
\cline{1-3}
Model & MAP & NDCG \\
\hline
BM25  & 0.1977 & 0.4334\\
QL & 0.1913 & 0.4325\\
\hline
\multicolumn{3}{c}{NON ENRICHED MODELS }\\
\hline
GloVe &  0.1951 &  0.4337 \\
GloVe.42B.300d &  0.1947 &  0.4308\\
GloVe.6B.100d &  0.1903 &  0.4291\\
GloVe.6B.200d &  0.1947 &  0.4308\\
GloVe.6B.300d &  0.1947 &  0.4308\\
GloVe.6B.50d &  0.1799 &  0.4119\\
GloVe-Twitter-100 &  0.1863 &  0.4218\\
GloVe-Twitter-200 &  0.1863 &  0.4218\\
GloVe-Twitter-25 &  0.1391 &  0.3488\\
GloVe-Twitter-50 &  0.1812 &  0.4147\\
w2v-GoogleNews-300 &  0.1947 &  0.4308\\
dict2vec-dim100 &  0.1995 &  0.4335\\
dict2vec-dim200 &  0.1959 &  0.4315\\
dict2vec-dim300 &  0.1957 &  0.4315\\
WordNet & 0.1977 & 0.4334 \\
\hline
\multicolumn{3}{c}{ENRICHED MODELS }\\
\hline
counterfit-GloVe &  0.1801 &  0.4027\\
GloVe-retrofitted &  0.1940 &  0.4264\\
paragram-counterfit &  0.1663 &  0.3906\\
paragram-74627 &  0.2005 &  0.4365\\
paragram-retrofitted &  0.1798 &  0.4012\\
paragram-retrofitted-74627 &  0.1798 &  0.4012\\
w2v-76427 &  0.1964 &  0.4318\\
w2v-counterfit-header &  0.1734 &  0.3991\\
w2v-retrofitted &  0.1967 &  0.4368\\
\hline
\multicolumn{3}{c}{AFFECT ENRICHED MODELS }\\
\hline
GloVe-affect &  0.1947 &  0.4308\\
counterfit-GloVe-affect &  0.1810 &  0.4044\\
GloVe-retrofitted-affect-555 &  \textbf{0.2021} &  \textbf{0.4421}\\
paragram-affect &  0.1977 &  0.4309\\
paragram-counterfit-affect &  0.1844 &  0.4094\\
w2v-affect &  0.1940 &  0.4305\\
w2v-counterfit-affect &  0.1762 &  0.4029\\
w2v-retrofitted-affect &  0.1971 &  0.4345\\
\hline
\end{tabular}
\caption{Query expansion experiments on NYT }
\label{tab:expansionNYT}
\end{table}

Table \ref{tab:rankingCACM} shows performance for Ranking experiments on the CACM dataset using different embeddings. 
We observe that usage of term embeddings generally causes steadily higher NDCG and MAP. In particular the paragaram embeddings models report the best results, with affect enriched paragram embeddings reporting both best NDCG and MAP, $0.02$ better than non affect enriched paragram embeddings results in both NDCG and MAP.

\begin{table}[H]
\begin{tabular}{llr}
\hline
\multicolumn{3}{c}{Ranking experiments on CACM} \\
\cline{1-3}
Model & NDCG & MAP \\
\hline
BM25 & 0.3805 & 0.1947\\
QL & 0.3621 & 0.2056\\
\hline
\multicolumn{3}{c}{NON ENRICHED MODELS }\\
\hline
GloVe.42B.300d & 0.3638 & 0.2007\\
GloVe.6B.100d & 0.4440 & 0.2722\\
GloVe.6B.200d & 0.4452 & 0.2732\\
GloVe.6B.300d & 0.4450 & 0.2730\\
GloVe.6B.50d & 0.4437 & 0.2720\\
GloVe-Twitter-100 & 0.5109 & 0.3260\\
GloVe-Twitter-200 & 0.5138 & 0.3292\\
GloVe-Twitter-25 & 0.5309 & 0.3217\\
GloVe-Twitter-50 & 0.4682 & 0.2715\\
w2v-GoogleNews-300 & 0.3697 & 0.1960\\
GloVe & 0.4483 & 0.2760\\
\hline
\multicolumn{3}{c}{ENRICHED MODELS }\\
\hline
counterfit-GloVe & 0.4563 & 0.2680\\
GloVe-retrofitted & 0.4507 & 0.2787\\
w2v-76427 & 0.4920 & 0.3033\\
w2v-counterfit-header & 0.4085 & 0.2225\\
w2v-retrofitted & 0.3993 & 0.2350\\
paragram-counterfit & 0.5675 & 0.3722\\
paragram-74627 & 0.5539 & 0.3541\\
paragram-retrofitted & 0.5263 & 0.3467\\
paragram-retrofitted-74627 & 0.5380 & 0.3633\\

\hline
\multicolumn{3}{c}{AFFECT ENRICHED MODELS }\\
\hline
counterfit-GloVe-affect & 0.4247 & 0.2383\\
GloVe-affect & 0.4326 & 0.2553\\
w2v-affect & 0.3900 & 0.2080\\
w2v-counterfit-affect & 0.3791 & 0.2006\\
w2v-retrofitted-affect & 0.3555 & 0.1986\\
paragram-affect & 0.5848 & 0.3986\\
paragram-counterfit-affect & \textbf{0.5860} & \textbf{0.3996}\\
\hline
\end{tabular}
\caption{Ranking experiments on CACM}
\label{tab:rankingCACM}
\end{table}

Table \ref{tab:expansionCACM} shows performance for query expansion experiments on the CACM dataset using different embeddings. 
We observe that usage of term embeddings generally causes steadily higher NDCG and MAP. While we expected best results with Aff2Vec models it turned out "vanilla" word2vec model trained on Google News corpus outperformed all the others in NDCG and MAP.
On the other hand the best performing enriched model is a retrofitted word2vec model whereas among affect enriched models the GloVe retrofitted one provides the best results.

\begin{table}[H]
\begin{tabular}{llr}
\hline
\multicolumn{3}{c}{Query expansion experiments on CACM} \\
\cline{1-3}
Model & NDCG & MAP \\
\hline
BM25 & 0.3805 & 0.1947\\
QL & 0.3621 & 0.2056\\
\hline
\multicolumn{3}{c}{NON ENRICHED MODELS }\\
\hline
WordNet & 0.4014 & 0.2146\\
GloVe.42B.300d & 0.4657 & 0.2701\\
GloVe.6B.100d & 0.4646 & 0.2635\\
GloVe.6B.200d & 0.4633 & 0.2631\\
GloVe.6B.300d & 0.4724 & 0.2707\\
GloVe.6B.50d & 0.4575 & 0.2588\\
GloVe-Twitter-100 & 0.4500 & 0.2576\\
GloVe-Twitter-200 & 0.4454 & 0.2524\\
GloVe-Twitter-25 & 0.4215 & 0.2373\\
GloVe-Twitter-50 & 0.4422 & 0.2528\\
w2v-GoogleNews-300 & \textbf{0.4824} & \textbf{0.2828}\\
GloVe & 0.4635 & 0.2685\\

\hline
\multicolumn{3}{c}{ENRICHED MODELS }\\
\hline
counterfit-GloVe & 0.4622 & 0.2661\\
GloVe-retrofitted & 0.4676 & 0.2723\\
w2v-76427 & 0.4366 & 0.2518\\
w2v-counterfit-header & 0.4557 & 0.2629\\
w2v-retrofitted & 0.4738 & 0.2816\\
paragram-counterfit & 0.4661 & 0.2716\\
paragram-74627 & 0.4626 & 0.2712\\
paragram-retrofitted & 0.4470 & 0.2636\\
paragram-retrofitted-74627 & 0.4486 & 0.2646\\

\hline
\multicolumn{3}{c}{AFFECT ENRICHED MODELS }\\
\hline
counterfit-GloVe-affect & 0.4622 & 0.2673\\
GloVe-affect & 0.4694 & 0.2734\\
GloVe-retrofitted-affect-555 & 0.4722 & 0.2799\\
w2v-affect & 0.4609 & 0.2643\\
w2v-counterfit-affect & 0.4579 & 0.2674\\
w2v-retrofitted-affect & 0.4667 & 0.2744\\
paragram-affect & 0.4426 & 0.2586\\
paragram-counterfit-affect & 0.4634 & 0.2723\\
\hline
\end{tabular}
\caption{Query expansion experiments on CACM}
\label{tab:expansionCACM}
\end{table}

\section{Conclusions}
We present extensive experiments to evaluate the impact of affect-enriched word embeddings for information retrieval over a news corpus, namely ranking and query expansion implemented using open-source toolkits. We show that using affect-enriched models shows a significant improvement for ranking against baseline/vanilla embeddings~(\~20\%) as well as other enriched embeddings~(\~2-10\%). In case of query expansion, improvement is observed for the NYT dataset but vanilla GloVe embeddings report highest values for the CACM dataset. We believe the semantic structure and vocabulary distribution of the CACM dataset results in this behavior. 
We plan to extend this work first towards understanding the role of semantic information in expansion tasks and then towards building fusion approaches leveraging enriched word vectors with standard IR baselines.

\bibliographystyle{alpha}
\bibliography{sample-bibliography}

\end{document}